# A precessing jet from an active galactic nucleus drives gas outflow from a disk galaxy

Justin A. Kader[1*], Vivian U[2,1*], Loreto Barcos-Muñoz[3,4], Marina Bianchin[1], Sean T. Linden[5], Yiqing Song[6,7], Gabriela Canalizo[8], Archana Aravindan[8], George C. Privon[3,4,9], Tanio Díaz-Santos[10,11], Christopher Hayward[12]†, Matthew A. Malkan[13], Lee Armus[2], Rosalie C. McGurk[14], Jeffrey A. Rich[15], Anne M. Medling[16], Sabrina Stierwalt[17], Claire E. Max[18], Aaron S. Evans[3,4], Christopher J. Agostino[19], Vassilis Charmandaris[10,11,20], Tianmu Gao[21,22], Justin H. Howell[2], Hanae Inami[23], Thomas S.-Y. Lai[2], Kirsten L. Larson[24], Christopher D. Martin[25], Mateusz Matuszewski[25], Joseph M. Mazzarella[2], James D. Neill[25], Nikolaus Z. Prusinski[25], Raymond Remigio[1], David B. Sanders[26], Jason Surace[2]

1. Department of Physics and Astronomy, University of California Irvine, Irvine, USA
2. IPAC, California Institute of Technology, Pasadena, USA
3. National Radio Astronomy Observatory, Charlottesville, USA
4. Department of Astronomy, University of Virginia, Charlottesville, USA
5. Department of Astronomy, University of Arizona, Tucson, SA
6. European Southern Observatory, Santiago, Chile
7. Joint Atacama Large Millimeter Array Observatory, Santiago, Chile
8. Department of Physics and Astronomy, University of California Riverside, Riverside, USA
9. Department of Astronomy, University of Florida, Gainesville, USA
10. Institute of Astrophysics, Foundation for Research and Technology-Hellas, Heraklion, Greece
11. School of Sciences, European University Cyprus, Engomi, Cyprus
12. Center for Computational Astrophysics, Flatiron Institute, New York, USA 13. Department of Physics and Astronomy, University of California, Los Angeles, USA
14. W. M. Keck Observatory, Kamuela, USA
15. The Observatories of the Carnegie Institution for Science, Pasadena, USA
16. Department of Physics and Astronomy and Ritter Astrophysical Research Center, University of Toledo, Toledo, USA
17. Physics Department, Occidental College, Los Angeles, USA
18. Department of Astronomy and Astrophysics, University of California Santa Cruz, Santa Cruz, USA
19. NPC Worldwide, LLC, Bloomington, USA
20. Department of Physics, University of Crete, Heraklion, Greece
21. Research School of Astronomy and Astrophysics, Australian National University, Weston Creek, Australia
22. Australian Research Council Centre of Excellence for All Sky Astrophysics in Three Dimensions, Canberra, Australia
23. Hiroshima Astrophysical Science Center, Hiroshima University, Hiroshima, Japan
24. Association of Universities for Research in Astronomy for the European Space Agency, Space Telescope Science Institute, Baltimore, USA
25. Cahill Center for Astronomy and Astrophysics, California Institute of Technology, Pasadena, USA
26. Institute for Astronomy, University of Hawaii at Manoa, Honolulu, USA
*Corresponding authors. Email: jukader@gmail.com, vivianu@ipac.caltech.edu
† Present address: J.P. Morgan Securities LLC, New York, USA

**To reproduce observed galaxy properties, cosmological simulations require that massive galaxies experience feedback from active galactic nuclei, which regulates star-formation within those galaxies. However, the energetics and timescales of these feedback processes are poorly constrained. We combine optical, infrared, sub-millimeter and radio observations of the active galaxy VV 340a, hosting a low-power jet launched from a supermassive black hole at its center. We find that the jet undergoes precession, with a period of $(8.2 \pm 5.5) \times 10^5$ years, and drives an outflow of gas at a rate of $19.4 \pm 7.85$ solar masses per year. The jet shocks the gas, producing highly ionized plasma extending several kiloparsecs from the nucleus. The outflow ejects sufficient gas from the galaxy to influence its star formation rate.**

Galaxies grow over time as dynamic entities. Theoretical models indicate that feedback from stars and the accretion discs around supermassive black holes influences the composition and physical properties of the gas and can potentially eject some of the gas out of the galaxy (*1*). An actively accreting supermassive black hole (an active galactic nucleus, AGN) can drive a sufficiently strong galactic outflow to eject most of the gas and therefore halt star formation within the host galaxy (*2*). An AGN phase can be triggered by a merger or other gravitational interaction between galaxies (*3, 4*). The details of the AGN feedback process are poorly constrained by observations.

AGN feedback regulates star formation through two primary modes. The ejective mode is typically observed in star-forming late-type (e.g., spiral) galaxies, where radiatively efficient AGN (those rapidly converting accreting matter into intense X-ray and UV radiation) drive outflows that expel gas from the host (*5, 6*). In contrast, the preventative mode is associated with quiescent, giant elliptical galaxies, driven by powerful, collimated jets from radio-quiet AGN (whose output is dominated by mechanical energy and strong radio emission) that heat the surrounding gas and prevent it from cooling and accreting (*7, 2*). Observations of jet-driven outflows in radio-quiet AGN (those without the large-scale radio emission typical of radio-loud sources) (*8-15*), including within spiral galaxy disks (*16-18*), have shown that non-radiative, mechanical feedback can sometimes be the dominant mode in star-forming galaxies, complicating the simple two-mode paradigm. Numerical simulations predict that low-power jets remain trapped in the host galaxy gas for longer than high-power jets; during this time, they generate turbulence, shocks, and entrain gas into outflows (*19, 20*). We searched for a galactic-scale, radio jet-driven outflow in a gas-rich spiral disk galaxy to test models of jet feedback.

**The VV 340a galaxy**
VV 340 (also known as UGC 09618 and Arp 302) is a pair of spiral galaxies in the early stages of merging, making their first approach towards each other (*21*). They have a redshift $z = 0.034$ (*22*) (equivalent to a luminosity distance of $D_L \sim 157$ megaparsecs) and the two galaxies are separated by 27.3 kiloparsecs (kpc) (*23*). In Hubble Space Telescope (*HST*) optical broadband imaging (*24, 25*), one galaxy is seen face-on (VV 340b) and the other is edge-on (VV 340a) (Fig. 1A). The total infrared (IR) luminosity is $L_{IR} = 5 \times 10^{11}$ solar luminosities ($L_\odot$), of which 90% originates in the northern component VV 340a, so it is classified as a luminous infrared galaxy (LIRG). Infrared observations of the nucleus of VV 340a revealed an emission line at 14.3 μm, identified as the highly ionized coronal line [Ne V] (*23*)—an emission line from ions with a high ionization potential (IP > 100 electron volts or eV), with this specific line having an IP of 97 eV. X-ray observations have shown a hard (2-7 keV) X-ray source also coincident with the nucleus, indicating that VV 340a contains an AGN that is obscured from view by dust (*23*). Near-IR observations of gas in the nuclear region show no shocked outflows of molecular gas (*26*). However, there is an X-ray nebula which extends outwards from the nucleus to several kpc above the galactic disk (Fig. S5A), which is a potential sign of a galactic-scale outflow (*27*). We selected VV 340a for additional observations because its edge-on orientation facilitates the robust measurement of outflow extent and morphology with minimal contamination from the galactic disk, a geometric advantage shared with the starburst galaxy Messier 82, where this view simplifies kinematic analysis (*28*).

**Extended coronal line emission**

We observed VV 340a with the Mid-Infrared Instrument (MIRI) on the James Webb Space Telescope (*JWST*), using its integral-field spectroscopy mode (*25*). These observations show a collimated extraplanar structure extending from the nucleus toward the northeast (NE) and southwest (SW) with emission in several emission lines), including the highly ionized [Ne V] coronal line discussed above. The continuum-subtracted [Ne V] emission is shown in Fig. 1B. We constructed velocity channel maps of the extended [Ne V] emission (Fig. 2) which show that the collimated gas has projected velocities of < 500 km s$^{-1}$ at a radial distance 1.8 kpc from the nucleus. The velocity is away from the observer in the NE and towards the observer in the SW. The emission morphologies and velocities are consistent with a symmetric bipolar ionized outflow from the nucleus. This highly collimated coronal line gas extends at least 3 kpc from the nucleus.

Models of AGN feedback predict that the hottest gas carries a substantial fraction of the kinetic energy of the outflow (*29, 30*), but such hot gas is only directly observable in X-rays. Our observation of coronal lines provides an indirect tracer of the hot gas. This gas could have been heated by shock fronts driven by the passage of radio jets (*14, 31*).

**Galactic-scale ionized nebulae**

We performed integral-field spectroscopic observations of VV 340a with the Keck Cosmic Web Imager (KCWI) on the Keck-II telescope (*25*), targeting the [O II] λλ3726,3729, Hβ λ4861, and [O III] λ5007 emission lines (where λ notation indicates the wavelength in angstroms, Å). Figure 1B shows the continuum-subtracted [O III] emission which forms two bright collimated nebulae extending away from the IR nucleus towards the NE and SW, extending 11 kpc and 15 kpc, respectively. These nebulae are aligned with the inner coronal line outflow visible in the *JWST* data (Fig. 1B). Higher resolution KCWI observations of the [O III] line in the inner 3 kpc of VV 340a show that the [O III]-emitting gas is tightly collimated there, sharing the same morphology as the [Ne V]-emitting coronal gas (*25*) (Fig. S5B). The KCWI observations indicate that the highly ionized, collimated outflow visible in the inner 3 kpc in the *JWST* data extends farther into the galactic halo.

The logarithmic line ratio log([O III]/Hβ) ranges from 0.56 to 1 along the nebulae (Fig. S3A). These values are inconsistent with stellar photoionization, indicating that the gas within the filament is either shock-ionized or photoionized by a hard radiation field from the AGN. Because the high-resolution [O III] data show that this gas follows a filamentary structure in the inner regions of the galaxy, aligned with the coronal line gas (Fig. S5A&B), these two gas phases may represent a stratified outflow, with different levels of exposure to a common ionizing source. There is an additional large nebula southeast (SE) of the nucleus, with a roughly circular projection and extending ~9 kpc above the galactic disk. Previous *HST* narrowband imaging of the galaxy (*32*) showed Hα arcs at the location of the eastern edge of the nebula. At this location the ratio log([O III]/Hβ) = 0.42, consistent with ionization by stellar radiation. We interpret the SE nebula as a supernova-driven outflow related to ongoing star formation in the disk.

**A radio jet undergoing precession**

We reanalyzed archival radio observations of VV 340a taken with the Karl G. Jansky Very Large Array (VLA) (*33, 25*). The resulting 6 GHz radio continuum image (Fig. 1B and Fig. S6A) shows a bipolar S-shaped jet aligned with the collimated filaments seen in both the mid-IR coronal lines and optical emission lines. The jets extend NE and SW in a straight line for ~2.7 kpc then abruptly but symmetrically change direction to align with the galaxy's minor axis, then extend for another ~1.7 kpc.

Symmetric S-shaped radio structures have previously been interpreted as due to jets undergoing precession (*34-36*), either due to the presence of binary supermassive black holes or accretion disk instabilities (*37*). We fitted models of precessing AGN jets (*38, 25*) to the VLA 6 GHz image, assuming a sub-relativistic plasma ejection velocity of $10^4$ km s$^{-1}$ (*35, 39, 40*). The best-fitting model parameters are a jet precession angle of $\phi = 14° \pm 8.8°$, a precession axis position angle (PA) of $\psi = 33° \pm 24°$, an inclination angle of $i = 71° \pm 9.8°$, and a precession period of $P = (8.2 \pm 5.5) \times 10^5$ yr; all uncertainties are 1σ. Fig. S1 shows the full parameter distributions from the iterative fitting procedure while Fig. S4 shows a schematic diagram of the jet geometry.

**Molecular gas reservoir**

We observed VV 340a with the Atacama Large Millimeter/submillimeter Array (ALMA), targeting the CO ($J = 2\rightarrow1$) emission line, where $J$ is the rotational quantum number (*25*). The total line intensity, velocity and velocity dispersion are shown in Fig. S8. The CO line velocity map (Fig. S8B) is consistent with disk-like rotation (~ 3 x 19 kpc in size) coincident with the optical and radio disks. There is no substantial molecular gas at the locations of the jet or [O III] nebulae (Fig. S8A).

We also examined archival ALMA observations of the CO ($J = 1\rightarrow0$) line (*25*). Combining both lines provides an estimate of the CO gas mass, which we then converted to all molecular gas (*25*). The total molecular disk gas mass is $(1.7 \pm 0.17) \times 10^{10}$ $M_\odot$, which is high for a local LIRG (*41*). We set an upper limit on molecular gas in the outflow of $< (2.2 \pm 0.5) \times 10^8$ $M_\odot$ (on each side of the nucleus), which corresponds to < 1.3% of the total molecular gas. The lack of molecular gas on scales < 6 kpc (the maximum recoverable scale of the ALMA observations) could be explained by the destruction of any outflowing molecular gas by the radio jet. This scenario is consistent with the presence of the ionized gas nebulae seen in the [O III] data.

**AGN properties**

We use the [Ne V] flux, measured within a circular aperture with a radius of 0.7 kpc centered on the IR nucleus, and previously published methods (*42*) to calculate the AGN bolometric luminosity $L_{AGN,bol} = (1.28 \pm 0.06) \times 10^{44}$ erg s$^{-1}$, equivalent to $(3.3 \pm 0.1) \times 10^{10} L_\odot$. This implies a black hole accretion rate of $\dot{M}_{BH} = (2.3 \pm 0.1) \times 10^{-2} M_\odot$ yr$^{-1}$, assuming 10% efficiency in the conversion of mass to energy, typical for the local Universe (*25, 43-45*). We confirmed the bolometric luminosity using the [O III] flux measured in the same aperture and an empirical bolometric correction (*46, 25*), which gives indistinguishable results.

We also estimated the AGN bolometric luminosity using a *Chandra* hard (2-7 keV) X-ray energy spectrum (*23*), finding $L_{AGN,bol} = 2.1^{+7.54}_{-0.58} \times 10^{43}$ erg s$^{-1}$, which is six times lower than indicated by the [Ne V] and [O III] data (*25*). We also estimated the unobscured hard X-ray luminosity using the ALMA data and a correlation between millimetre and X-ray flux (*47, 25*), which provide an $L_{AGN,bol}$ that is consistent with the [Ne V] and [O III] data. We interpret this discrepancy as due to attenuation of the X-rays from the AGN by an optically thick obscuring medium along the line of sight (LOS), such as a dusty torus.

To determine the black hole mass ($M_{BH}$), we fitted a model to determine the stellar velocity dispersion ($\sigma_*$) in the inner 5 kpc of the galaxy, finding $\sigma_* = 127 \pm 16$ km s$^{-1}$, By combining this with an empirical $M_{BH}$–$\sigma_*$ relation (*48*), we estimate $M_{BH} = (2.9 \pm 0.36) \times 10^7 M_\odot$. The bolometric luminosity and black hole mass indicate it is accreting at $(3.7 \pm 0.5)$% the Eddington rate (the maximum theoretical accretion rate at which outward radiation pressure exactly balances the inward gravitational force). This value is consistent with either advection-dominated accretion flows (ADAFs, where hot gas falls into the black hole rapidly and most energy is trapped, often leading to powerful jets) or radiatively efficient accretion (where gas is cool and slow, releasing most of its energy as light, which drives strong radiative winds) (*49-52*). The outflow could therefore be driven by either an AGN wind or kinetic energy input from the jet, or a combination of the two, perhaps switching between them over time. Thermal winds, which are powered by the AGN accretion disk radiation field, generate nearly isotropic ionized outflows with wide opening angles (*53*), which we do not observe in VV 340a. The collimated structures of the coronal and low-ionization emission lines in the outflow, and their alignment with the radio jet, lead us to conclude that the outflow is being driven by the kinetic energy of the jet.

**Modeling the outflow**
We fitted a model of a precessing radio jet-driven outflow to the morphological and kinematic information measured from the [O III] and [Ne V] data (Fig. 3). This model assumes that as the jet precesses, it entrains material that flows outward into a hollow biconical structure. Because the precession axis is inclined with respect to the galactic disk, the jet's path length through the disk varies over the precession cycle. Consequently, the jet entrains material preferentially during the periods of longest path length, leading to a non-uniform appearance in the ionized bicone (Fig. 3). This model is consistent with the observed alignment between the jet and the [O III] filaments, and the extended [O III] nebula eastward of the NE filament (Fig. 4C).

The 0.5-2 keV soft X-ray image (*23*) shows X-ray emission coincident the radio jet and an elongation aligned with the ionized filaments (Fig. S5A). This indicates a mechanical outflow, powered by a jet, which generates shocks as it propagates through the gas within the galaxy. In the NE part of the outflow, there are double-peaked emission line profiles in both [Ne V] (Fig. 2B and

Fig. 3E) and in [O III] (Fig. 3D and Fig. S5C), a signature of a hollow bicone structure in which the near and far walls have differing LOS velocities. Although double-peaked emission lines are not observed in the ionized gas to the SW of the nucleus, there is instead a fork-like structure in the [Ne V] emission (Fig. 2A) which we interpret as the limb-brightened walls (meaning the projected path length through the hollow gas structure is maximized at the edges) of a bicone.

The extent to which gas fills the walls of the bicone depends on the jet power and the amount of gas entrained in it at each phase of the precession cycle. If these are not constant over the precession cycle, we expect the outflowing gas to form an incomplete bicone structure. The [O III]/Hβ ratios measured in the SE and NW nebulae, the appearance of Hα arcs on the edges of the SE nebula, and the extension of the soft band X-ray emission to the SE of the nucleus (Fig. S5A) could indicate that these nebulae are gas from supernova-driven winds. Alternatively, we cannot rule out the possibility that the precession axis of the jet was formerly oriented in the SE-NW direction, depositing gas in these nebulae before the jet re-oriented to the current NE-SW direction.

We compared the gas kinematics in the precessing jet-driven outflow model to the [Ne V] and [O III] observations. To facilitate this comparison, we assumed a bicone geometry defined by the jet precession and a simple velocity scaling. The latter assumes that gas accelerates outward along the bicone until it reaches a maximum velocity $v_{max}$ at a turnover radius of $r_t$, then decelerates. The projected LOS velocities that ultimately were compared with the observations were produced by summing the LOS velocities from the biconical outflow model with the LOS velocities predicted by a separate galactic bulge-disk-halo dynamical model (*25*) and then extracted along two synthetic slits matching those slits used to measure the velocities of the [Ne V] and [O III] lines (Fig. S2). The position-velocity (P-V) diagrams (Fig. 3D and Fig. 3E) demonstrate that while the kinematics along the main disk of the galaxy can be described entirely by the galactic bulge-disk-halo dynamical model, the addition of the biconical outflow model is required in order to reproduce the kinematics of the off-disk extended gas components. The model was further tested by comparing gas kinematics with those observed in the 15.6 μm [Ne III] emission line and kinematics of the [O III] line observed at higher spectral and angular resolution (Fig. S5C and Fig. S5D).

We cannot rule out the possibility that the S-shaped morphology of the radio jet is caused by symmetric deflection of the jet as it impacts dense gas on either side of the galaxy, changing the path of the expanding radio-emitting plasma, which could also explain the observed emission line kinematics (*51, 54, 55*). However, the extension of [O III] gas and X-ray emission beyond the deflection points leads us to favor the precessing jet scenario, with emission powered by previous precession cycles of the jet having faded as the jet plasma cooled and diffused.

**Galactic feedback**
We estimate the mass outflow rate from the model using $v_{max}$, the bicone cross-sectional area, and a direct measurement of the electron density from the [Ne V] 14 μm / 24 μm line flux ratio. This calculation utilizes a measurement of the electron temperature from the [O III] 4363 Å / (4959 + 5007 Å) line flux ratio and assumes a typical coronal line region filling factor (*56, 25*). We find an outflow rate of $\dot{M}_{out} = 19.4 \pm 7.85\ M_\odot\ yr^{-1}$ (*25*). The primary systematic uncertainty in this result stems from the adopted, unconstrained filling factor, which is uncertain by over an order of magnitude. However, by adopting this standard assumption, the derived value and its reported statistical uncertainty facilitate direct comparison with the results of most published studies that utilize similar assumptions for the filling factor. An alternative calculation using the electron density estimated from the [O II] observations gives a similar outflow rate of $\dot{M}_{\text{out}} = 13.8 \pm 5.94$ $M_\odot$ yr$^{-1}$ (*25*).

Adopting a star formation rate for this galaxy of $48 \pm 5.1$ $M_\odot$ yr$^{-1}$, previously derived from near-infrared imaging (*32*), we calculate the mass-loading factor (the ratio of the mass outflow rate to the galactic star formation rate) of the outflow $\eta = 0.4 \pm 0.26$, with similarly high uncertainty. The mass outflow rate is not sufficient to halt star formation in the galaxy. We place a lower limit on the gas depletion timescale, the amount of time for all cold molecular gas in the disk to form stars or be ejected from the galactic disk. For this calculation, we assume all the ionized gas in the outflow was initially cold molecular gas in the galactic disk, before being ionized and entrained in the outflow by the radio jet. To calculate this timescale, we used a standard galactic evolution framework (the leaky box model) that accounts for gas flowing out of the galaxy as well as new gas being returned to the disk from dying stars (the stellar return fraction is assumed to be 0.425) (*57*). Given the total molecular gas mass of $M_{H_2} = (1.7 \pm 0.17) \times 10^{10}$ $M_\odot$ (*41*), the gas depletion timescale is $t_d = (3.6 \pm 0.85) \times 10^8$ yr. Compared to the depletion timescale derived from star formation alone, $t_{d,SF} = (6.2 \pm 1.0) \times 10^8$ yr, the outflow accelerates depletion by shortening the timescale by $\Delta t_d = (2.5 \pm 0.9) \times 10^8$ yr. A correlated Monte Carlo analysis confirms this is a statistically significant difference at the $2.85\sigma$ level. If star-forming gas continues to be funneled toward the AGN for $\sim 3 \times 10^8$ yr, and if the jet remains active for that time, then it will have a substantial effect on the star formation of the galaxy.

Using the mass outflow rate and the turbulent velocity measured from the [Ne V] data ($326 \pm 24.4$ km s$^{-1}$), we calculate that the kinetic power in the outflow is $(1.0 \pm 0.40) \times 10^{43}$ erg s$^{-1}$ (*25*). VV 340a has a total luminosity density at 6 GHz of $8.9 \times 10^{22}$ W Hz$^{-1}$, and with a jet kinetic power of $(3.9 \pm 2.25) \times 10^{43}$ erg s$^{-1}$, measured using archival 1.49 GHz VLA observations (*58, 25*), it is classified as a radio-quiet source (*59*) with a low-power jet. However, some radio-quiet AGNs do produce radio jets that affect the evolution of their host galaxy. The ratio of the outflow kinetic power to the jet kinetic power is $0.26 \pm 0.18$, which is at the high end of the range (0.01 to 0.3) predicted by simulations of radio-mechanical feedback and inferred from observations of radio-loud early type galaxies (*11, 53*). Although we cannot rule out a starburst contribution to the origin of the extended nebulae, the collimated appearance and high emission line ratios we measured for the extraplanar gas indicate that the radio jet is the dominant driving mechanism. For the jet to be dominant, it must be converting a higher fraction of its power into the kinetic power of the outflow than is typical for AGNs. This interpretation is consistent with simulations of jet-driven outflows, which have predicted that low-power jets have a larger impact on the ISM of their host galaxy than high power jets do (*19*). This is because low-power jets propagate more slowly, which enhances the coupling between the jet and the ISM (*19*), especially when the jets have a low inclination angle from the galactic disk (*20*).

The ALMA CO observations show there is no substantial cold molecular gas in the outflow (< 1.3% of the total mass of molecular gas in the disk) (*25*). It is therefore possible that some of the kinetic energy of the jet is consumed by conversion of the molecular gas to ionized gas, contributing to the large outflow-to-jet kinetic power ratio. We do not observe more than one cycle of the precessing jet, so it is possible that the jet has only recently restarted, and the extended [O III] filaments were deposited in a prior activity phase. Although a secular (unrelated to merger activity) origin for the AGN activity in VV 340a cannot be ruled out, gravitational perturbations during the early stages of the merger with VV 340b could drive sufficient gas towards the AGN to trigger accretion at a few percent of the Eddington rate, which then produces the jet-driven outflow.

**Summary of AGN activity in VV 340a**
From our multiwavelength dataset and modelling, we conclude that the local star-forming LIRG VV 340a has an ongoing kiloparsec-scale outflow that is ejecting $19.4 \pm 7.85$ $M_\odot$ yr$^{-1}$ of ionized gas. The central part of the outflow consists of elongated collimated nebulae, which are sufficiently highly ionized to be observed in coronal emission lines, extending at least 3 kpc from the nucleus. The outflow then transitions to lower-ionization gas, observed as [O III] nebulae that extend a further ~10 kpc. No molecular gas is observed in the outflow. Radio continuum imaging shows an S-shaped radio jet that extends for several kiloparsecs and is aligned with the highly ionized gas filaments. X-ray emission is observed along the jet and ionized gas filaments, which is consistent with the presence of shocks and indicates that the jet is the driving mechanism of the outflow. The S-shape indicates that the jet is precessing as it propagates outward. The jet converts ~25% of its kinetic power into driving the outflow, although this value does not account for power consumed by the destruction of molecules. We conclude that VV 340a has a low-power radio jet which shock-ionizes and ejects gas as it propagates outward from the supermassive black hole and precesses on scales of several kpc.

**Acknowledgments**
This work is based on observations made with the NASA/ESA/CSA James Webb Space Telescope. The data were obtained from the Mikulski Archive for Space Telescopes (MAST) at the Space Telescope Science Institute, which is operated by the Association of Universities for Research in Astronomy, Inc., under NASA contract NAS 5-03127 for JWST.

Some of the data were obtained at Keck Observatory, which is a private 501(c)3 non-profit organization operated as a scientific partnership among the California Institute of Technology, the University of California, and the National Aeronautics and Space Administration. The Observatory was made possible by the generous financial support of the W. M. Keck Foundation.

This research made use of the Keck Observatory Archive (KOA), which is operated by the W. M. Keck Observatory and the NASA Exoplanet Science Institute (NExScI), under contract with the National Aeronautics and Space Administration.

The authors wish to recognize and acknowledge the very important cultural role and reverence that the summit of Maunakea has always had within the Native Hawaiian community. We are most fortunate to have the opportunity to conduct observations from this mountain.

The National Radio Astronomy Observatory is a facility of the National Science Foundation operated under cooperative agreement by Associated Universities, Inc.

ALMA is a partnership of ESO (representing its member states), NSF (USA) and NINS (Japan), together with NRC (Canada), MOST and ASIAA (Taiwan), and KASI (Republic of Korea), in cooperation with the Republic of Chile. The Joint ALMA Observatory is operated by ESO, AUI/NRAO and NAOJ.

This research has made use of data obtained from the *Chandra* Data Archive provided by the *Chandra* X-ray Center (CXC).

This work was performed in part at the Aspen Center for Physics, which is supported by National Science Foundation grant #PHY-2210452.

**Funding**
J.A.K. and V.U acknowledge funding support from STScI programs #JWST-GO-01717.001-A and #HST-GO-17285.001-A, NASA Astrophysics Data Analysis Program (ADAP) grant #80NSSC20K0450, NASA Astrophysics Decadal Survey Precursor Science (ADSPS) grant #80NSSC25K0169, National Science Foundation (NSF) Astronomy and Astrophysics Research Grant (AAG) #AST-2536603, and the Everglow Charitable Foundation.

H.I. acknowledges support from JSPS KAKENHI grant No. JP21H01129 and the Ito Foundation for Promotion of Science.

A.M.M. acknowledges support from the NASA Astrophysics Data Analysis Program (ADAP) grant number 80NSSC23K0750, from NSF AAG grant #2009416 and NSF CAREER grant #2239807, and from the Research Corporation for Science Advancement (RCSA) through the Cottrell Scholars Award CS-CSA-2024-092.

T.G. acknowledges support from Australian Research Council Discovery Project DP210101945.

**Author contributions**
V.U conceived the Keck observations, V.U, J.A.K., M.B., T.G., R.R., and R.M. performed Keck observations. J.A.K. led the data reduction and analysis of the Keck data. V.U, J.A.R., and K.L.L. conceived and planned the *JWST* observations. M.B. and K.L.L. aided in the reduction of the *JWST* data. J.A.K. led the analysis of the *JWST* data. L.B.M. and Y.S. analyzed the VLA data. L.B.M. led the analysis of the ALMA data. T.G. aided with the calculations of the X-ray luminosity. G.P. contributed to the calculation of the black hole mass. T.D.S., A.M.M., J.A.R., and S.S. contributed to the interpretation of the [O III] line kinematics. C.H. contributed to the discussion of star formation suppression. M.M., L.A., A.S.E., V.C., J.H.H., H.I., T.L., J.M.M., D.S., and J.S. contributed to the interpretation of the *JWST* data, determined the AGN properties and interpreted the [Ne V] kinematics. J.A.K. led the modeling of the radio jet with contributions from C.J.P. and C.E.M.. J.A.K. wrote the manuscript, with contributions from V.U. All co-authors commented on the manuscript and discussed the results.

**Competing interests**
We declare no competing interests.

**Data and materials availability**
The *JWST* data are available at MAST https://mast.stsci.edu/portal/Mashup/Clients/Mast/Portal.html by selecting 'MAST Observations by Proposal ID' and entering '01717'. The Keck medium- and low-resolution data are available at the Keck Observatory Archive https://koa.ipac.caltech.edu using 'more search options' and entering U213 for 'Program ID', 2022A for 'Semester', and U for 'Principal investigator'. The Keck high-resolution data can be found using the same method but entering E327 for 'Program ID'. The VLA data are available at the National Radio Astronomy Observatory Archive https://data.nrao.edu/portal/ under programs AL746 (6 GHz data) and AC205 (1.5 GHz data). The ALMA data are available at the ALMA Science Archive https://almascience.nrao.edu/aq/ under project codes 2017.1.01235.S and 2023.1.00499.S. The *Chandra* data are available in the Chandra X-ray Center Archive https://cda.harvard.edu/chaser/ under proposal number 08700551. The *HST* observations are available at the Hubble Legacy Archive https://hla.stsci.edu/ using the advanced search with 'vv340' in the main search box and 10592 in the 'Proposal ID' field.

**Supplementary Materials**
Materials and Methods
Figs. S1 to S8
References (*60-95*)

**Notice**
This is the author's version of the work. It is posted here by permission of the AAAS for personal use, not for redistribution. The definitive version was published in Science on January 8, 2026, DOI: 10.1126/science.adp8989; https://www.science.org/doi/10.1126/science.adp8989.

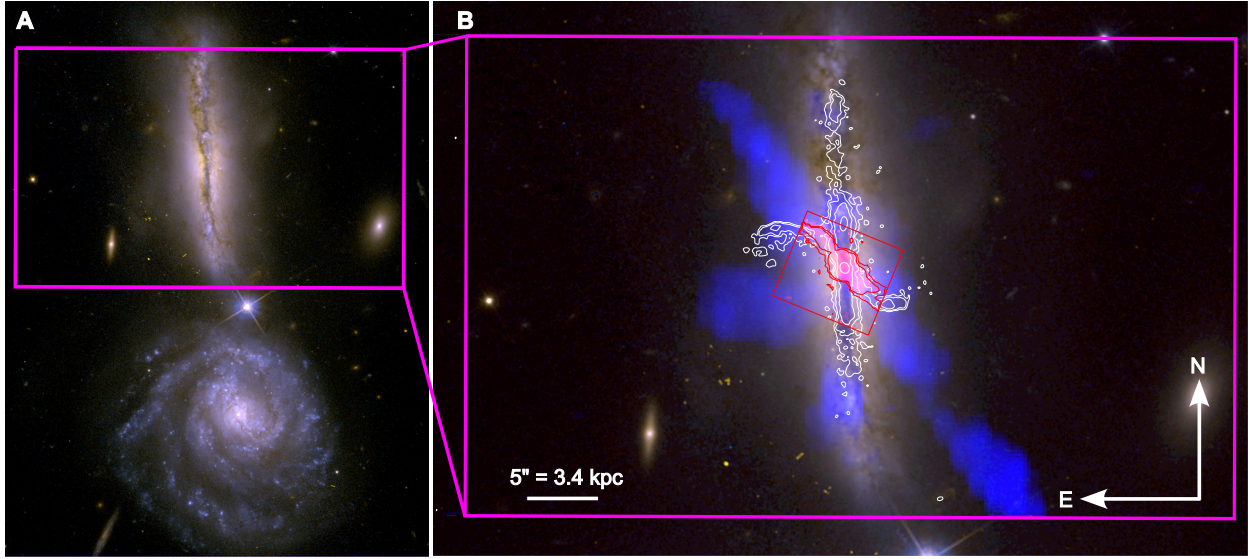

**Fig. 1: Broadband optical and multi-wavelength images of VV 340.** (**A**) Optical HST image of the interacting galaxies VV 340a (above) and VV 340b (below) (*91*). (**B**) Multi-wavelength composite image of VV 340a. The red and magenta boxes indicate the coverage of the *JWST* MIRI and Keck KCWI observations, respectively. The [Ne V] 14 μm emission from *JWST* MIRI is in red, the broadband optical HST image in green, and the [O III] emission from Keck KCWI is in blue. White contours show the VLA 6 GHz continuum map at (3, 5, 10, and 30) ×10$^{-4}$ Jy beam$^{-1}$ (1 Jy = 10$^{-23}$ erg s$^{-1}$ cm$^{-2}$ Å$^{-1}$). Red contours trace the [Ne V] 14 μm emission at (0.14 and 0.26) ×10$^{-16}$ erg s$^{-1}$ cm$^{-2}$ Å$^{-1}$. A scale bar of 5″ (equivalent to 3.4 kpc at the distance of VV 340) and a compass are labelled. The collimated coronal line emission is aligned with the radio jet, and the [O III] filaments continue outward. On the SW side, the ionized outflow extends 15 kpc from the central AGN.

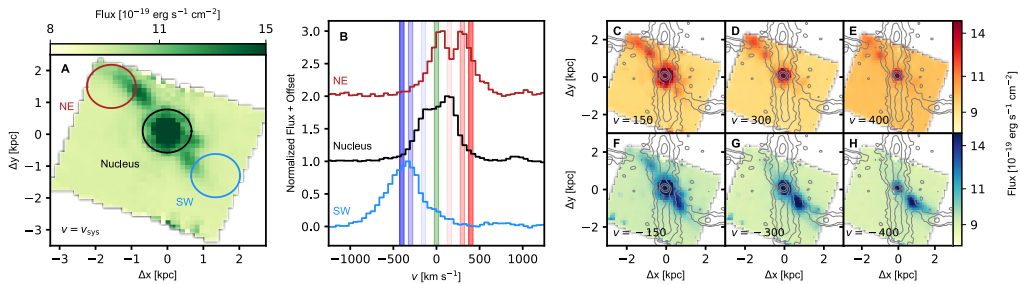

**Fig. 2: Channel maps and spatial distribution of the [Ne V] emission.** (**A**) Spatial distribution of the [Ne V] 14 μm coronal emission line flux from the MIRI observations, integrated within a spectral window corresponding to a LOS radial velocity $v = v_{sys} \pm 25$ km s$^{-1}$, where $v_{sys}$ is the systemic radial of VV 340a, which represents the velocity of the entire galaxy relative to the observer. ΔRA and ΔDec are the projected physical offset from the center of the nucleus in the [Ne V] map. Circles represent three 1″ extraction regions used to produce panel B. NE is a location northeast of the nucleus Nucleus is the IR nucleus, and SW is a location southwest of the nucleus. (**B**) Redshift-corrected and normalized spectra of the [Ne V] line in the three regions marked in panel A, vertically offset by unity for display. Overlain colored bands are velocity channels at 0, ±150, ±300, and ±400 km s$^{-1}$, each with widths of 50 km s$^{-1}$. (**C-H**) Channel maps of the [Ne V] line at the same velocities as marked in panel B, labelled in each panel. The velocity channel is indicated to the bottom right of each panel in km s$^{-1}$. The VLA 6 GHz continuum emission is represented by contours, as in Fig. 1, colored grey here. The high-velocity gas emitting [Ne V] forms a symmetric bidirectional outflow that extends for ~3 kpc. The double-peaked spectrum in the NE indicates a hollow cone structure.

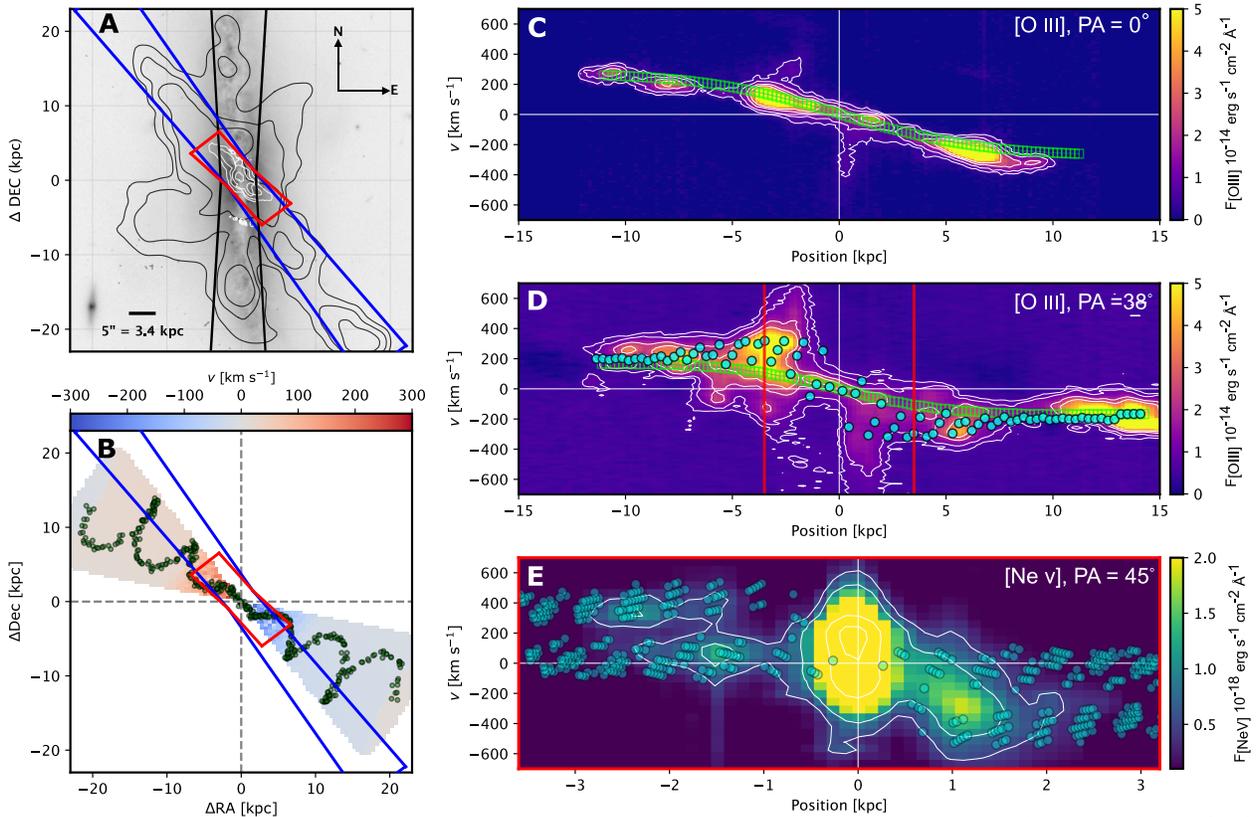

**Fig. 3: Kinematic analysis of the [O III] and [Ne V] observations**. (**A**) Inverted grayscale optical *HST* image (*24*) of VV 340a overlain with contours indicating the velocity-integrated line flux for [Ne V] (white) and [O III] (black). Boxes indicate the synthetic slits used for panels C-E, for [O III] along the galactic disk (PA = 0°; black) along the extended [O III] nebulae (PA = 38°; blue), and along the [Ne V] nebulae (PA = 45°; red). (**B**) The model helical precessing jet (green circles) and the model hollow biconical outflow, both projected onto the plane of the sky. The red and blue shading on the bicone represent the LOS radial velocity (*v*) of the outflowing gas from the model. The synthetic slits aligned with the outflow shown in Panel A are overlain. (**C** and **D**) position–velocity (P–V) diagrams for [O III], using the synthetic slits placed along the galaxy major axis (PA = 0°) and along the extended nebulae (PA = 38°) respectively. In both panels, color indicates emission line flux density. White contours are isophotes at 0.5 ×, 1 ×, 2 ×, and 3 ×10$^{-14}$ erg s$^{-1}$ cm$^{-}$

$^2$ Å$^{-1}$. Open green boxes are the LOS velocity bulge-disk-halo dynamical model projected along the disk major axis (panel C) and the filaments (panel D). Cyan filled circles in panel D are the LOS velocities of the biconical outflow model; red lines indicate the region shown in panel E. (**E**) the P–V data for the [Ne V] gas, measured in the synthetic slit at PA = 45°. The maximum velocity in the model is labelled. Color indicates [Ne V] line flux density, with white contours at 0.05 ×, 0.1 ×, 0.5 ×, and 1 × 10$^{-18}$ erg s$^{-1}$ cm$^{-2}$ Å$^{-1}$. Cyan circles are the same as in panel D. The [Ne V] and [O III] P-V diagrams are matched by a combination of gas following the biconical outflow model and galactic rotation.

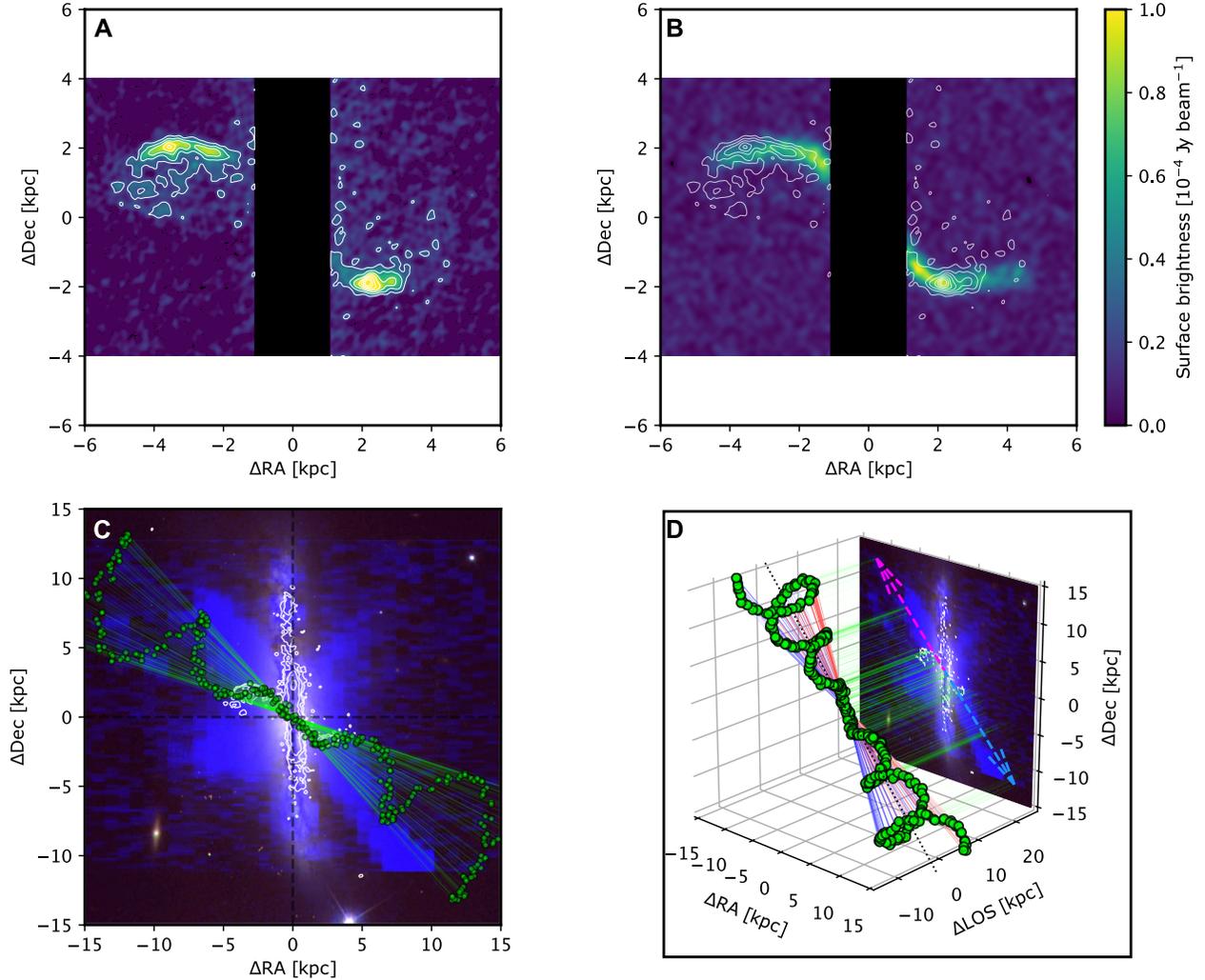

**Fig. 4: Radio observations compared to the precessing jet model.** (**A**) White contours show the observed 6 GHz radio continuum emission at 2 ×, 5 ×, 8 ×, 10 ×, and 13 × 10$^{-5}$ Jy beam$^{-1}$. Color indicates 6 GHz surface brightness. (**B**) synthetic image of the best-fitting precessing jet model generated by projecting the model onto the plane of the sky and adding Gaussian noise. Overlain white contours are the same as in panel A. The black area in panels A and B has been masked out to exclude the galactic disk. (**C**) Green circles show the precessing jet model, simulated over 3 precession periods. The background is a false color *HST* image of the galaxy; the blue overlay is [O III] line flux. White contours are the same as in panels A and B, except without masking out the galactic disk. Green straight lines are the plasma ejection velocity vectors. (**D**) Same as panel C,

but the model is on a three-dimensional orthographic projection and plasma ejection vectors are color-coded blue and red according to their angle against the ΔLOS direction. Green lines are projections of the three-dimensional jet positions onto the plane of the sky (the two-dimensional image). The blue and pink dashed arrows indicate the current on-sky projected orientation of the jet in its precession cycle. The model jet reproduces the positions of the radio continuum emission. The morphologies of the extended [O III] nebulae approximately follow the current ejection velocity vectors.

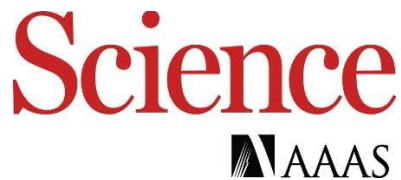

Supplementary Materials for

**A precessing jet from an active galactic nucleus drives gas outflow from a disk galaxy**

Justin A. Kader*, Vivian U*, Loreto Barcos-Muñoz, Marina Bianchin, Sean T. Linden, Yiqing Song, Gabriela Canalizo, Archana Aravindan, George C. Privon, Tanio Díaz-Santos, Christopher Hayward, Matthew A. Malkan, Lee Armus, Rosalie C. McGurk, Jeffrey A. Rich, Anne M. Medling, Sabrina Stierwalt, Claire E. Max, Aaron S. Evans, Christopher J. Agostino, Vassilis Charmandaris, Tianmu Gao, Justin H. Howell, Hanae Inami, Thomas Lai, Kirsten L. Larson, Christopher D. Martin, Mateusz Matuszewski, Joseph M. Mazzarella, James D. Neill, Nikolaus Z. Prusinski, Raymond Remigio, David Sanders, and Jason Surace

*Corresponding authors: Justin A. Kader, jukader@gmail.com, Vivian U, vivianu@ipac.caltech.edu

**The PDF file includes:**

Materials and Methods
Figs. S1 to S8

**Materials and Methods**

Cosmology

Throughout this work we assume a concordance flat dark-energy cold dark matter cosmological model with a Hubble constant $H_0$ = 69.3 km s$^{-1}$ Megaparsec$^{-1}$, dark energy density parameter $\Omega_\Lambda$ = 0.714, and matter density parameter $\Omega_m$ = 0.286, based on the 9-year Wilkinson microwave anisotropy probe cosmology (*60*).

*JWST* observations and analysis
Mid-infrared integral-field spectroscopic observations of VV 340a were conducted with *JWST* MIRI (*61, 62*) in medium resolution spectroscopy (MRS) mode on Universal Time (UT) 2023 March 23-24 as part of general observers cycle 1 program GO-1717. Exposures across the short, medium, and long sub-bands of each of the four wavelength channels resulted in observations that cover the full 4.9 to 28.8 µm range. The fast readout mode with an additional reset between integrations (FASTR1) was chosen to optimize the dynamic range expected in the observations. We used the standard 4-point dither pattern in each sub-band observation to optimize the sampling in the dispersion and cross-dispersion directions. The integration time was 444 s per exposure. Because the source extends beyond the instrument field of view (FOV), we obtained a dedicated background exposure with the same observational configuration in each of the three sub-bands.

The uncalibrated frames were downloaded from the Barbara A. Mikulski archive for space telescopes (MAST) and reduced using the *JWST* science calibration pipeline (*63*) version 1.11.0 in MRS batch mode, which includes steps to apply detector-level corrections, wavelength calibration, flux calibration, background subtraction, and to co-add dither frames. The final calibrated cubes had a FOV of (3.″2×3.″7), (4″.0 ×4.″8), (5.″2×6.″2), and (6.″6×7.″7), and a spatial resolution of 0.″196, 0.″196, 0.″245, and 0.″273 for channels 1, 2, 3 and 4, respectively. We analyzed the collisionally excited forbidden emission line of the Ne$^{4+}$ ion at a rest wavelength ($\lambda_{rest}$)= 14.322 µm, (hereafter [Ne v]), which was observed at a redshifted wavelength of ~14.8 µm in MRS Channel 3. The spectral resolution of the instrument at this wavelength is $6 \times 10^{-3}$ µm (equivalent to 120 km s$^{-1}$), corresponding to an instrumental resolving power of ~2500. The same observations included detections of the [Ne v] 24 µm ($\lambda_{rest}$ = 24.318 µm) and [Ne III] 15.6 µm ($\lambda_{rest}$ = 15.555 µm) lines.

We estimated the AGN bolometric luminosity ($L_{AGN,bol}$) using empirical correlations between [Ne v] luminosity and several AGN bolometric indicators (*42*). The flux of the [Ne v] line ($F_{[Ne\,v]}$) was measured in a cylindrical aperture centered on the bright IR nucleus, with a radius set to 1″, which is twice the full width at half maximum (FWHM) of the MIRI point spread function at 14 µm. The flux was $F_{[Ne\,v]} = (1.15 \pm 0.06) \times 10^{-14}$ erg s$^{-1}$ cm$^2$. The statistical uncertainty on the flux was ~5 per cent, derived from the square root of the diagonal of the parameter covariance matrix computed by the Levenberg-Marquardt algorithm used when fitting Gaussian profiles to the emission lines. At the adopted co-moving distance of 157 Mpc, we find $L_{AGN,bol} = (1.28 \pm 0.06) \times$ 10 erg s$^{-1}$ = $(3.3 \pm 0.1) \times 10^{10}\, L_\odot$ from the [Ne v] line. Assuming an accretion efficiency of $\eta$ = 0.1, which is typical for a geometrically thin accretion disk (*43-45*), the black hole accretion rate is $\dot{M}_{Edd} = \eta c^2 L_{AGN,bol} = (2.3 \pm 0.1) \times 10^{-2}\, M_\odot$ yr$^{-1}$.

Fig. 2 shows velocity channel maps of the [Ne v] coronal emission line observed with MIRI. The systemic velocity channel (Fig. 2A) shows the bright IR nucleus plus collimated nebulae (filaments) directed radially outward on symmetrically opposite sides, to the NE and SW. The coronal line gas to the NE and SW has LOS velocities spanning from systemic to approximately ± 500 km s$^{-1}$, with the NE side redshifted and the SW side blueshifted. The [Ne v] emission shows two filaments to the SW that form a forked morphology, while in the NE, the [Ne v] lines are double-peaked. The P-V diagram of the [Ne v] line emission was constructed using synthetic two-dimensional slits applied to the MIRI data cube using the software QFitsView (64). The continuum emission was sampled in the line-free 12000-4500 km s$^{-1}$ channel and then subtracted from the P–V diagrams. The slit was placed along the filamentary NE–SW structures at PA = 45º, with a central width of 3″ and tapering outward with an opening angle of 3º. The slit position is indicated in Fig. 3A. The resulting P–V diagram is shown in Fig. 3E. Fig. 2 shows velocity channel maps of the [Ne v] 14 μm coronal emission line observed with MIRI. The systemic velocity channel (Fig. 2A) shows the bright IR nucleus plus collimated filaments directed radially outward on symmetrically opposite sides, to the NE and SW. The coronal line gas to the NE and SW has LOS velocities spanning from systemic to ~+500 km s$^{-1}$ and ~-500 km s$^{-1}$, respectively, with the NE side redshifted and the SW side blueshifted. The [Ne v] emission shows two filaments to the SW that form a forked morphology, while in the NE the [Ne v] lines are double-peaked. The P-V diagram of the [Ne v] line emission was constructed using synthetic two-dimensional slits applied to the MIRI data cube using the software QFitsView (61). The continuum emission was sampled in the line-free 12000-4500 km s$^{-1}$ channel and then subtracted from the P–V diagrams. The slit was placed along the filamentary NE–SW structures at PA = 45º, with a central width of 3″ and tapering outward with an opening angle of 3º. The slit position is indicated in Fig. 3A. The resulting P–V diagram is shown in Fig. 3E.

Keck observations and data analysis
Optical integral-field spectroscopic observations of VV 340a were obtained with KCWI (65) mounted at the right Nasmyth focal station of the Keck II telescope. Observations were made with the instrument configured with the blue high-resolution grating with the small slicer (BH-S mode), and the blue medium and blue low-resolution gratings with the large slicer (BM-L and BL-L modes). The resulting spectral resolving powers were $R$ ~ 4500, 2000, and 900, and the instantaneous spectral coverages were Δλ = 4960 to 5340, 3650 to 4550, and 3500 to 5500 Å, respectively. The BH-S mode had a FOV of 8″ × 20″ with slit width-limited spatial sampling (0.″3 pixel$^{-1}$ sampling rate along the 0.″3-wide slit, i.e., square pixels). The BM-L and BL-L modes had a FOV of 33″ × 20.″4 with slit width-limited spatial sampling (0.″3 pixel$^{-1}$ sampling rate along the 0.″7- and 1.″4-wide slits, respectively). The BM-L and BL-L KCWI observations were obtained on UT 2022 May 23 under clear sky conditions (seeing FWHM ~ 0.″9) as part of Keck observing program 2022A_U213. The BH-S observations were obtained on UT 2023 July 16 under clear skies as part of Keck observing program 2023B_E327. For BM-L and BL-L, we employed a 5-point dithering pattern centered on VV 340a, arranged in a 2 × 2 grid plus a central, overlapping pointing, for continuous coverage across a 60″ × 36″ area, plus one dedicated background sky pointing. For BH-S, there was a single pointing toward the nucleus of VV 340a. Integration times were 1200, 1200, and 600 s, for the BH-S, BM-L and BL-L configurations, respectively, with the same exposure times for science, dark, and background exposures. Basic reduction and wavelength solutions for the spectroscopic data were performed using the data reduction pipeline KCWI_DRP version 1.1 (66). We measured an instrumental FWHM of 1, 2 and 4 Å in the BH-S, BM-L, and BL-L configurations, respectively.

Several large cylindrical apertures with radius $r$ = 3″ were placed at the locations shown in Fig. S3A to investigate the ionization mechanisms of the gas. The [O III]/Hβ ratio is used to identify whether gas is ionized by radiation from stars, AGN, or shocks (67-70). The disk apertures have log([O III]/Hβ) = 0.25, -0.09, -0.12, at the nucleus, and at positions separated by 10″ (= 6.8 kpc) to the north and 11″ (= 7.5 kpc) south, respectively. Toward the SE nebula and the NW nebula, we find higher line ratios of log([O III]/Hβ) = 0.42 and 0.50, respectively. We find higher line ratios in the two filaments: log([O III]/Hβ) = 1 for the aperture placed 10″ from the nucleus along the NE filament, and log([O III]/Hβ) = 0.84 and 1.03 for two positions along the SW filament, 9″ (=6.1 kpc) and 20″ (= 13.6 kpc) from the nucleus. The SE nebula appears to emerge orthogonally to the disk, and the previously observed Hα arcs (32) might be portions of a bubble which has burst, so we posit that this gas is part of a kpc-scale supernova-driven outflow or galactic fountain (28, 71, 72). The [O III]/Hβ line ratios measured along the filaments are inconsistent with stellar photoionization, but could be explained by ionization by either a hard radiation field from the AGN or by shock ionization.

P-V diagrams of the [O III] line emission were constructed using the same method as for the [Ne v] line. The continuum emission was sampled in the 1200-1000 km s$^{-1}$ channel and then subtracted from the P–V diagrams. The first slit was placed along the galactic main disk, with a central width of 2″ and tapering outward with an opening angle of 3°. The second slit follows the filamentary NE–SW structures at PA = 38°, with a central width of 3″ and tapering outward with an opening angle of 3°. The slit positions are indicated in Fig. 3A and the P–V diagrams are shown in Fig. 3C-D.

We fitted a model of a composite bulge-disk-halo galactic potential to the [O III] velocity data using the Galpy software (73, 74). The bulge was modeled as a power-law density spherical potential with an exponential cut-off, with an inner power-law index of 1.8 and a cut-off radius of 0.1 kpc. The disk was modeled as a Miyamoto-Nagai potential (75) with a disk scale length of 5 kpc and a scale height of 5 kpc. The dark matter halo was modeled as a Navarro-Frenk-White potential (76) with a scale radius of 2 kpc. The relative normalizations of the bulge, disk, and halo potentials were set to 0.001, 1, and 0.002, respectively, before taking their sum as the model galactic potential. The model disk was projected to an inclination angle of 70° (24). The model fitting was performed using Markwardt-Levenberg least-squares minimization (77) using the residuals between the [O III] line velocity map and the model LOS velocity field. The model LOS velocities projected along the disk (PA = 0°) and filament (PA = 38°) directions are shown in Fig. 3C-D.

We derived an alternative measurement of $L_{\rm AGN,bol}$ using the [O III] line flux in a 1″ aperture centered on the nucleus of VV 340a using the KCWI BL-L data. We find an [O III] flux $F_{\rm [OIII]}$ = (3 ± 0.1) × 10$^{-16}$ erg s$^{-1}$ cm$^{-2}$. Adopting an empirical AGN bolometric correction (39), we calculate $L_{\rm AGN,bol}$ = (1.1 ± 0.21) × 10$^{44}$ erg s$^{-1}$ from the [O III] measurement, which is consistent with the value derived from [Ne v] above.

The black hole mass $M_{\rm BH}$ was estimated using the black hole mass-stellar velocity dispersion ($M_{\rm BH}$–$\sigma_*$) scaling relation of nearby AGNs (48). Penalized pixel fitting (78,79) was used to fit the 3600 to 4400 Å spectral region (containing the Ca II H+K stellar absorption features) with a model stellar line of sight velocity distribution (LOSVD), using a stellar template library (80). The best-fitting LOSVD for the central 5″ ×5″ (r ~ 5 kpc) indicates an instrumental dispersion-corrected $\sigma_*$ = 127 ± 16 km s$^{-1}$. This implies an estimated $M_{\rm BH}$ = (3.0 ± 5.21) × 10$^7$ $M_\odot$. From this mass, we calculated the Eddington accretion rate $\dot{M}_{\rm Edd}$ = 4πG/ηcκ = $6.5^{+11.5}_{-4.16}$ × 10$^{-1}$ $M_\odot$yr$^{-1}$, where we assume the standard electron-scattering opacity for fully ionized gas κ = 0.34 cm$^2$ g$^{-1}$ (81). This is about 27×$\dot{M}_{\rm acc}$, or equivalently $\dot{M}_{\rm acc}$ / $\dot{M}_{\rm Edd}$ = 0.035. We estimated an Eddington luminosity of $L_{\rm Edd}$ = $3.7^{+6.56}_{-2.38}$ × 10$^{45}$ erg s$^{-1}$ = $1.0^{+1.71}_{-6.23}$ × 10$^{12}$ $L_\odot$, so the Eddington ratio $\lambda_{\rm Edd}$ = $3^{+6.0}_{-2.2}$ × 10$^{-2}$.

VLA data analysis

To investigate the radio continuum emission we used archival VLA C band (4 to 8 GHz) observations in its A and B array configurations (*33*). We re-reduced the archival C band images following previous methods (*82*). The A and B configuration images were combined into a composite image with a 1.′3 × 1.′3 FOV covering VV 340a and part of VV 340S, with a synthesized circular beam FWHM of 0.″4. We also utilized archival VLA L band 1.5 GHz observations taken in the C array configuration to produce a radio continuum image with a 15″ beam (*58, 83*). Both radio continuum images are shown in Fig. S6.

Cylindrical apertures with radii 2.″7 and 2.″0 (Fig. S6) were used to measure the flux of the eastern and western jets respectively, yielding flux densities of 12.4 ± 1.27 mJy and 1.69 ± 0.170 mJy at 1.5 and 6 GHz for the eastern jet, and 8.85 ± 0.940 and 1.14 ± 0.114 mJy for the western jet. These values indicate a spectral index of -1.4 between 1.5 and 6 GHz, which is consistent with synchrotron-dominated emission. The 1.5 GHz flux densities were used to estimate the radio jet kinetic powers, because this image is more sensitive to the diffuse lower-level emission from the jet than the 6 GHz data is. Our calculation used $P_{jet}$-$P_{radio}$ scaling relations calibrated at 1.5 GHz based on observations of X-ray cavities in giant galaxies (*84, 85*). Depending on which scaling relation was used, we find the jet kinetic power was $(0.87 \pm 15.4) \times 10^{43}$ erg s$^{-1}$ for the eastern jet and $(0.68 \pm 11.96) \times 10^{43}$ erg s$^{-1}$ for the western jet (*84*), or $(2.20 \pm 30.31) \times 10^{43}$ erg s$^{-1}$ for the eastern jet and $(2.00 \pm 27.53) \, 10^{43}$ erg s$^{-1}$ for the western jet (*85*). In both cases, there is large (factor of ~10) uncertainty which is dominated by the scatter in the empirically calibrated scaling relations.

The S-shaped morphology of the 6 GHz emission is consistent with predictions from models of precessing radio jets (*37, 38*). Starburst or AGN wind shocks can accelerate relativistic electrons producing synchrotron emission (*86*), however we prefer the precessing jet scenario due to the high degree of symmetry on both sides of the galaxy. We interpret the radio continuum emission as cyclo-synchrotron emission from electrons with initially relativistic speeds that decelerate due to interactions with the interstellar medium (ISM).

To estimate the precession period of the jet, we compared the VLA image with models of ballistic helical precessing jets (*38*). The 3-D jet model parameters include: $\beta$, the plasma ejection velocity $v$ normalized to the speed of light $c$; the inclination angle $i$ of the jet with respect to the LOS; the precession angle $\phi$; the precession period $P$; and the precession axis position angle $\psi$. We generate a synthetic image of the jet by projecting the jet trace (composed of model jet plasma particles propagating along ballistic trajectories) onto the plane of the sky, construct a 2-D histogram of the plasma particles in each (RA, Dec) cell, and adding Gaussian noise to match the thickness of the jet in the VLA image. We fitted the synthetic image of the model jet to the VLA 6 GHz image using a linear least squares approach to minimize the likelihood function, which is taken to be the collapsed 1-D vector of squared residuals between the observed and synthetic jet images. We fixed the initial plasma velocity to $v = 10^4$ km s$^{-1}$ ($\beta = 0.033$), because this is a typical value for radio jets of this power (*35, 39, 40*). The other parameters were free to vary. The best-fitting model gives a jet precession axis PA of $\psi = 33° \pm 23.6°$, a precession angle of $\phi = 14° \pm 8.8°$, an inclination of $i = 71° \pm 9.8°$, and a precession period of $P = (8.2 \pm 5.48) \times 10^5$ yr. The 6 GHz image and the best-fitting projected 3-D model jet are shown in Fig. 4A-B. The 1-D parameter distributions are shown in Fig. S1.

ALMA observations and analysis

ALMA observations of VV 340a in the CO ($J = 2\rightarrow1$) [hereafter CO (2-1)] spectral line and ~1.3 mm continuum were obtained on UT 2023 December 8 (using configuration C-6, 44 antennas with maximum baselines of 2.5 km) and on 2024 April 27 and 28, (configuration C-3, 41 and 43 antennas with maximum baselines of 500 m) as part of project 2023.1.00499.S. The ALMA spectral line setup had four 1.875 GHz (~2400 km s$^{-1}$) wide spectral windows centered at 223.227 GHz (covering CO (2-1)), 224.922 GHz, 245.606 GHz, and 247.356 GHz, each with a spectral resolution of 1.129 MHz (1.5 km s$^{-1}$). We imaged the CO (2-1) line emission using the continuum-subtracted calibrated measurement provided by the observatory, which was processed through the ALMA calibration pipeline (*87, 88*). To match the VLA and *JWST* observations, we produced primary beam corrected CO (2-1) line cubes with a beam FWHM of 0.″4, pixel size of 0.″05, and spectral resolution of 6.4 km s$^{-1}$. We produced moment 0 (integrated intensity), moment 1 (intensity-weighted mean velocity), and moment 2 (velocity dispersion) maps across 133 channels (~840 km s$^{-1}$). We applied sigma-clipping to each image at 5σ, using a mask from the non-primary beam corrected cube. The resulting moment 0, 1, and 2 maps are shown in Fig. S6.

All the CO emission is concentrated in the disk with no substantial emission seen in the outflow. From the unclipped moment 0 map (root mean square velocity of 0.25 Jy km s$^{-1}$) we measure the total flux density of the line using a rectangular aperture 4″ wide and 24″ long at PA = 0°, covering the full extent of the molecular disk. The measured CO (2-1) flux density for the disk is 486 ± 49 Jy km s$^{-1}$, equivalent to $L_{CO(2-1)} = (7.1 \pm 0.7) \times 10^9$ K km s$^{-1}$ pc$^2$ using previous methods (*89*).

To convert this luminosity to mass we require $r_{21}$, the ratio of CO 2-1 to CO ($J = 1\rightarrow0$) [hereafter CO (1-0)]. We used archival ALMA archival CO (1-0) observations for VV 340 (project code 2017.1.01235.S), reduced in the same way as the CO 2-1 data. We produced a moment 0 map matching the velocity width and range of the CO (2-1) moment map described above (rms 0.24 Jy km s$^{-1}$). Using the same disk aperture, we estimate a CO (1-0) flux density of 159 ± 9 Jy km s$^{-1}$. This indicates an integrated $r_{21} = 0.75 \pm 0.09$. For the $r_{21}$ calculation, we lowered the spatial resolution of the CO (2-1) cube to ~0.″8, to match the CO 1-0 observations. Similar $r_{21}$ values have previously been reported local ultra-luminous infrared galaxies ((U)LIRGs), although for VV 340a it is lower than the median (*89*). Using the measured $r_{21}$ ratio and a conversion factor from CO(1-0) to H$_2$ of 1.8 $M_\odot$ (K km s$^{-1}$ pc$^{-2}$)$^{-1}$ [derived from local (U)LIRGs (*90*)], we derive a total molecular gas mass for the disk of $(1.70 \pm 0.17) \times 10^{10}$ $M_\odot$. Using the 0.″4 CO (2-1) moment 0 map and an aperture size of 1″ × 4″ for each lobe, we derive an upper limit for molecular gas in the outflow of < 1.3 per cent of the molecular mass of the disk.

We use the continuum image at 231.35 GHz (1.3 mm) delivered by the observatory, which has angular resolution 0.″19 × 0.″15, to measure the continuum peak flux density. We measure a peak flux density at the position of the core of 0.24 ± 0.06 mJy beam$^{-1}$, which corresponds to a luminosity of $1.65 \times 10^{39}$ erg s$^{-1}$. Assuming this mm flux is coronal emission from the central AGN, we use empirical mm-X-ray correlations for nearby AGN [(*47*), their table 1] to derive $L_{AGN,bol} = 4 \times 10^{44}$ erg s$^{-1}$ from the mm continuum.

*Chandra* data analysis

*Chandra* advanced CCD imaging spectrometer (ACIS) observations of VV 340a (*23*) consist of X-ray images in the full (0.4-7 keV), soft (0.4-2 keV) and hard (2-7 keV) bands, and an energy spectrum. The Chandra ACIS soft band (0.5 to 2 keV) X-ray image was smoothed with a Gaussian kernel with $r_{smooth} = 3$ pixels and $\sigma_{smooth} = 1.5$ pixels. The resulting image (Fig. S7) shows X-ray emission aligned and overlapping with the outflow, indicating thermal X-ray emission which we interpret as due to gas shocked by a jet (*86*).

The *Chandra* ACIS hard X-ray (2-10 keV) image was used to derived an alternative AGN bolometric luminosity measurement from. We modeled the spectrum from the central 1″ region as a power law continuum, plus a Gaussian for the broad Fe Kα feature at 6.4 keV. The low quality of the spectrum in this aperture makes it unsuitable for a direct measurement of the column density of the intrinsic absorption toward the AGN, so we adopt a value of the column density $N_H = 3.7 \times 10^{23}$ cm$^{-2}$, typical for local (U)LIRGs at early merger stages (*91*). The resulting intrinsic hard X-ray luminosity is $L_X = 1.35^{+3.46}_{-0.97} \times 10^{42}$ erg s$^{-1}$. Adopting an empirically-derived X-ray bolometric correction for AGNs (*92*), we find $L_{AGN,bol} = 2.1^{+7.54}_{-0.58} \times 10^{43}$ erg s$^{-1}$ from the X-ray data. This value is an order of magnitude smaller than those derived above from the optical, mid-IR, and mm measurements. We therefore regard it as an underestimate; it is more likely that the hard X-rays are attenuated by an obscuring medium with a higher column density than we assumed.

Outflow model
As the radio jet propagates outward from the AGN, it couples with the ISM of the galaxy, driving shocks and entraining gas in an outflow that moves outward radially from the AGN. As the jet precesses, the outflowing gas eventually forms a hollow bicone. At any given moment in time, the jet has a narrow interaction cross section with the ISM. The ISM is clumpy and its density varies strongly as a function of disk scale height. Therefore outflowing gas might not be distributed uniformly on the walls of the bicone.

To test our hypothesis that the structures seen in [Ne V] and [O III] are the incomplete walls of a large-scale biconical outflow, driven by a precessing jet, we compared their kinematics with predictions from an idealized biconical outflow model (*56, 93*). The hollow bicone is defined by the PA and inclination of its axis and the opening angles of its inner and outer walls, $\phi_{in}$ and $\phi_{out}$. De-projected velocities along the bicone walls were modeled using a velocity law of the form $v_r(r \leq r_t) = k_1 r$ and $v_r(r > r_t) = v_{max} - k_2 r$, where $v_r(r)$ is the deprojected radial velocity of the gas (viewed from the AGN), which is a function of radius $r$; $k_1$ and $k_2$ are the rates of linear acceleration and deceleration; and $r_t$ is the turnover radius where the maximum velocity $v_{max}$ is reached and deceleration begins.

Predictions from the biconical outflow model were compared with the measured kinematics of the [Ne V] and [O III] gas in two synthetic long slits placed along the directions of maximum elongation of the gas (PA = 45º for [Ne V] and PA = 38º for [O III]), shown in Fig. 3A-B. Fig. 3D and Fig. 3E show the P–V data extracted along these slits for both the observations and the total model in Fig. 3D. The total model is the sum of the velocities from both the outflow model and the galactic dynamical model in each position bin. There is good agreement between the data and model when the bicone has an orientation and inner opening angle set equal to the jet precession axis orientation and precession angle, an outer opening angle of 20º, and a velocity law with $k_1 = 1000$ s$^{-1}$ and $k_2 = 100$ s$^{-1}$, giving $v_{max} = 1.2 \times 10^3$ km s$^{-1}$ at $r_t = 1.25$ kpc.

We calculated the mass outflow rate of gas traced by the [Ne V] 14 μm coronal line using the biconical outflow model $\dot{M}_{out} = 2m_p n_e v_{max} A f$ (*56*), where $m_p$ is the proton mass, $n_e$ is the electron density of the gas, $v_{max}$ is from the bicone model, $A$ is the cross-sectional area of the bicone base (on one side) measured at $r_t$, and $f$ is the volume filling factor. The electron density was estimated directly from the [Ne V] 14 μm / 24 μm line flux ratio measured in two 1″ cylindrical apertures, centered on the collimated [Ne V] nebulae to the NE and SW (Fig. 2). We find ratios of $0.97 \pm 0.094$ and $1.00 \pm 0.079$, respectively. The theoretical line emissivity software PyNeb (*94, 95*) was employed to determine the electron density from the [Ne V] 14,24 μm line ratio and an estimate of the electron temperature, which we obtained from the [O III] auroral-to-nebular emission line ratio $R_{\text{[O III]}} = 0.02 \pm 0.008$ in the NE aperture, since the [O III] λ4363 auroral line was not detected in the SW. PyNeb indicates that the electron temperature and density in the outflow are $T_e = (1.83 \pm 0.291) \times 10^4$ K and $n_e = (1.07 \pm 0.408) \times 10^3$ cm$^{-3}$, respectively. The annular cross-sectional area of the bicone at $r_t$ is $A = 2.9 \times 10^5$ pc$^2$, calculated from the model parameters above, and $v_{\max} = 1.2 \times 10^3$ km s$^{-1}$. Assuming a filling factor of 0.001 (*56*), the ionized gas mass outflow rate $\dot{M}_{\text{out}} = 19.4 \pm 7.85$ $M_\odot$ y$^{-1}$. In a 1″-radius cylindrical aperture placed at $r_t$, the average turbulent velocity in the NE and SW apertures, measured from the broad component of the [Ne V] line, was $\sigma_{\text{turb}} = 326 \pm 24.4$ km s$^{-1}$, leading to an outflow kinetic power $\dot{E}_{\text{out}} = \tfrac{1}{2}\dot{M}_{\text{out}}(v_{\max}^2 + \sigma_{\text{turb}}^2) = (1.0 \pm 0.40) \times 10^{43}$ erg s$^{-1}$. The density measurement was repeated in the NE aperture using the [O II] 3726,3729 Å ratio observed in the BM-L data cube. We find a ratio of $R_{\text{[O II]}} = 1.06 \pm 0.127$, corresponding to an electron density of $n_e = (7.6 \pm 3.10) \times 10^2$ cm$^{-3}$ and $T_e = (1.78 \pm 0.300) \times 10^4$ K. Using this density, the mass outflow rate and outflow kinetic power are $\dot{M}_{\text{out}} = 13.8 \pm 5.94$ $M_\odot$ y$^{-1}$ and $\dot{E}_{\text{out}} = (7.2 \pm 3.55) \times 10^{42}$ erg s$^{-1}$. All of the outflow rates calculated here are upper limits, because they assume the outflowing gas is uniformly distributed on the bicone surface, contrary to the observations.

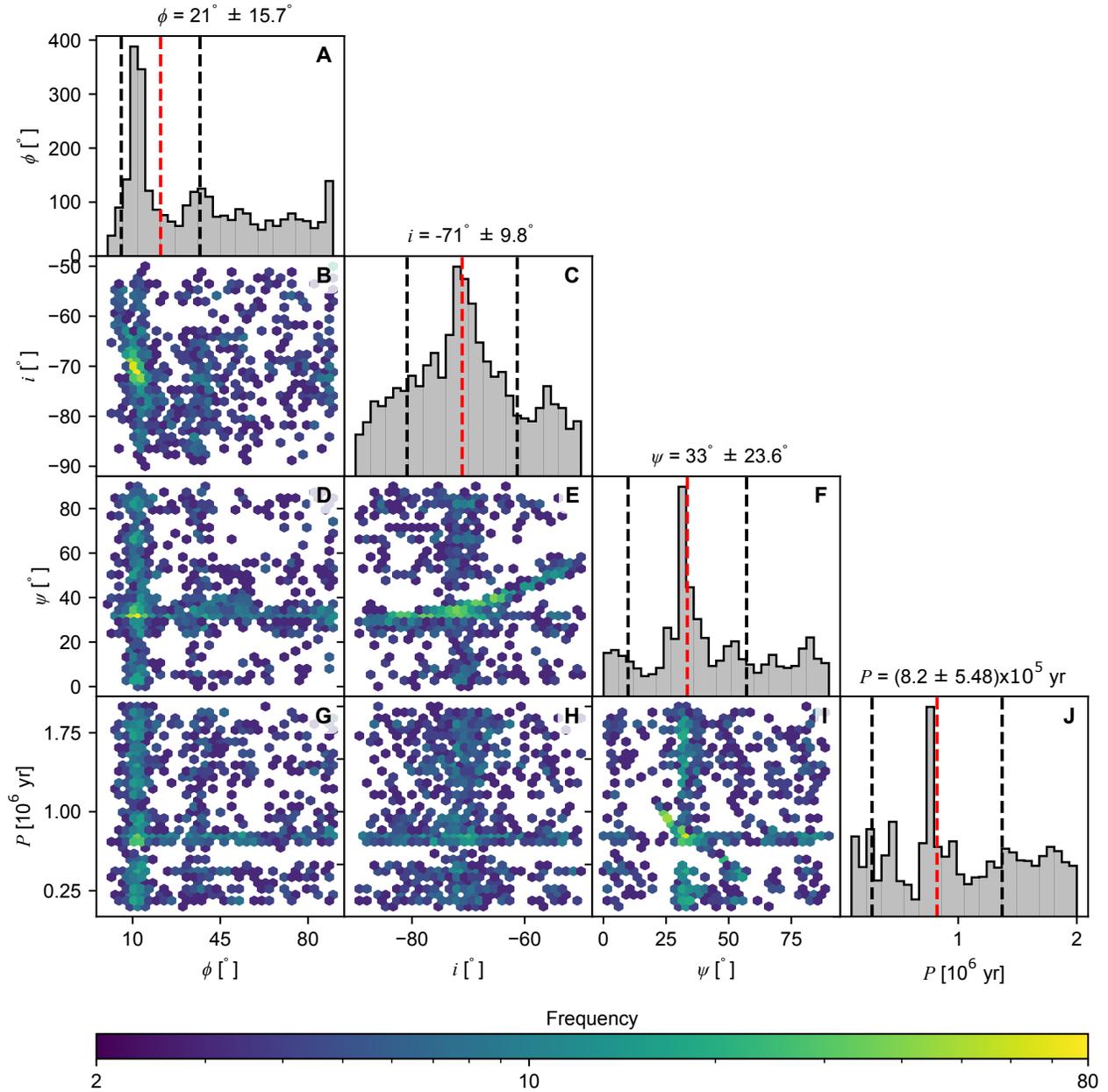

**Fig. S1.**
**Distributions of the helical precessing jet model parameters.** Each panel on the diagonal shows the 1-D distribution of a single model parameter fitted to the VLA 6 GHz image over 1000 least-squares minimization iterations. The 1-D distributions are overlain with the mean (blue dotted line) and median (red dashed line) computed after a single sigma clip (at 1σ), and the RMS standard deviation envelope is shown as the black dashed lines. Labeled at the bottom of the histograms are the median ± 1σ values of the distributions. Off diagonal panels show binned 2-D histograms of the joint distributions for each pair of parameters; color indicates the number of times a parameter pair was returned as a best fit. There is covariance between the jet PA ($\psi$) and the inclination angle ($i$), but not between the other parameters.

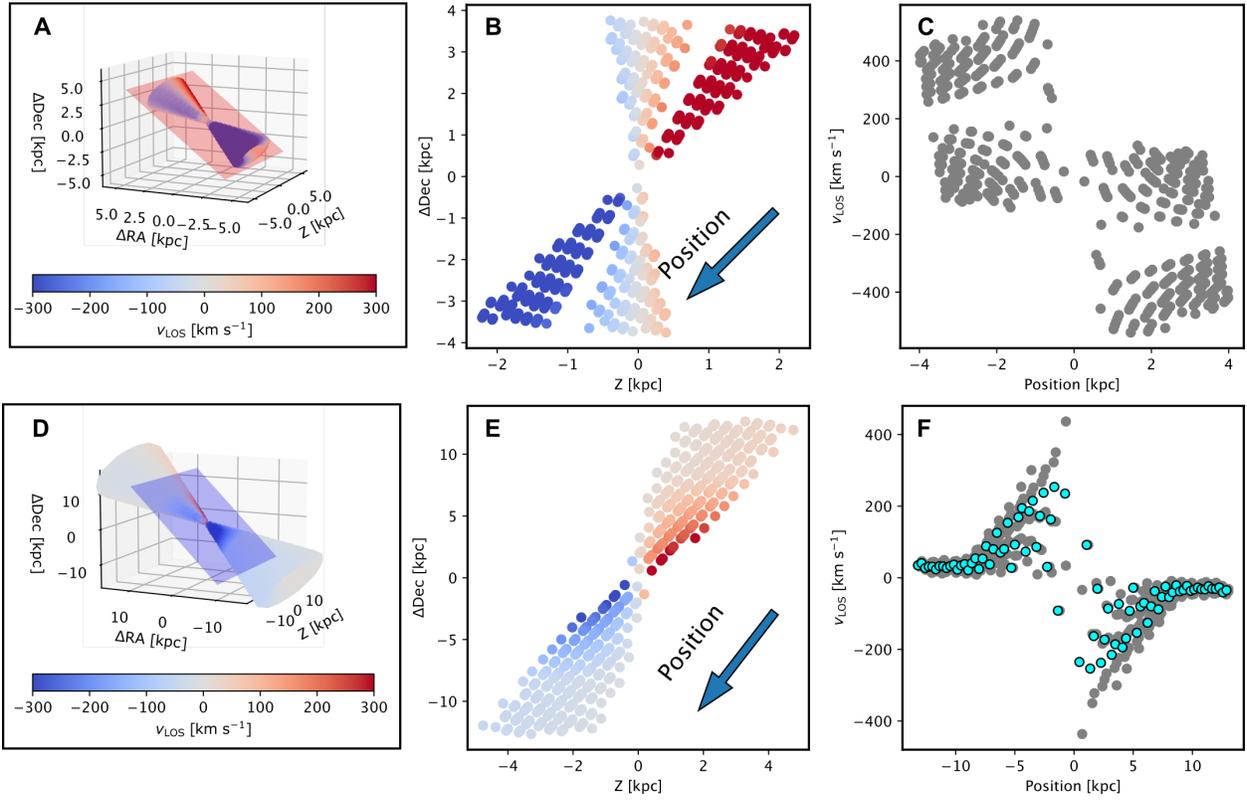

**Fig. S2.**
**Schematic illustration of P–V predictions from the biconical outflow model.** Panels A and D show the 3-D hollow bicone model colored according to the LOS component of the deprojected outflow velocity. The red and blue planes represent the synthetic slits matching the orientation and length of the slits used to extract P–V data for the [Ne v] (PA = 45º) and [O iii] (PA = 38º) gas, respectively. Panels B and E show a biconical cross section along the planes shown in Panels A and D. The filled colored circles represent the LOS component of the deprojected outflow velocity evaluated at each of the grid positions that comprise the bicone model. The arrow indicates the orientation of the synthetic slit used to extract the cross section. Panels C and F are P–V diagrams showing the LOS velocities as a function of position along the slits, mean velocities computed in position bins are shown in panel F (cyan filled circles). These are compared with the [O iii] and [Ne v] observations in Figs. 3D and 3E.

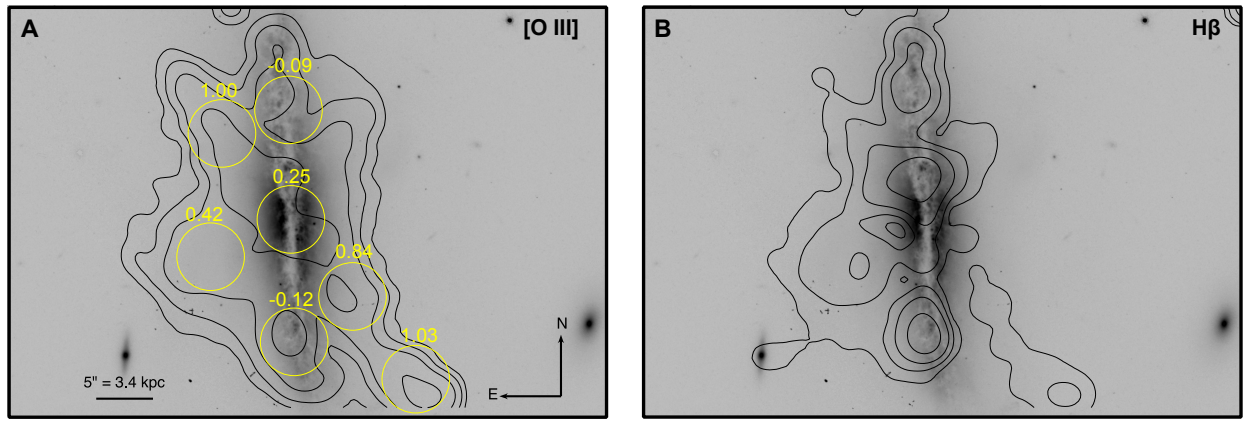

**Fig. S3**
**Broadband optical, [O III] and Hβ emission line maps of VV 340a.** (**A**) Grayscale *HST* broadband optical images of VV 340a, overlaid with black contours of the [O III] emission line flux. Contours are drawn at the 0.025, 0.1, 0.3, and $1\times10^{-16}$ erg s$^{-1}$ cm$^{-2}$ levels. Yellow circles are 3″ radius apertures used to measure the log([O III]/Hβ) ratio, with measured values indicated above each aperture. (**B**) Same as panel A but with contours of the Hβ flux.

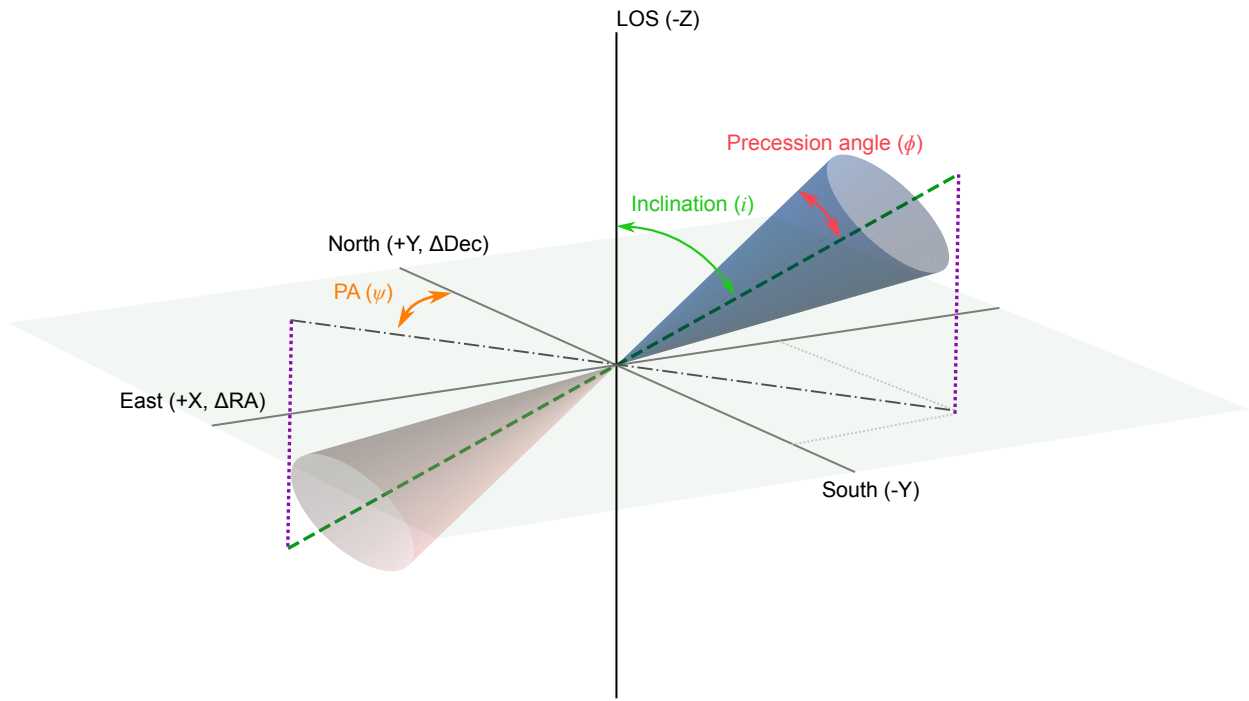

**Fig. S4**
**Schematic diagram of the precessing jet model.** The coordinate system is defined by the ΔRA and ΔDec axes which are the offsets from the galactic center in the East-West and North-South directions, respectively, measured in units of kpc, and the Z axis which is the offset from the galactic center along the LOS measured in kpc (all solid grey lines). The sky plane (semi-transparent grey) is spanned by the ΔRA and ΔDec axes at Z = 0. The diagram spans ±20 kpc in all three axes. The red and blue hollow cones represent the receding and approaching sides of the biconical outflow model and the dashed green line is the precession axis. The precession axis PA ($\psi$) is the angle between the positive ΔDec (North) axis and the sky plane projection of the precession axis. The inclination of the precession axis ($i$) is the angle between the negative LOS axis and the precession axis. The precession angle ($\phi$) is the angle between the precession axis and the cone surface.

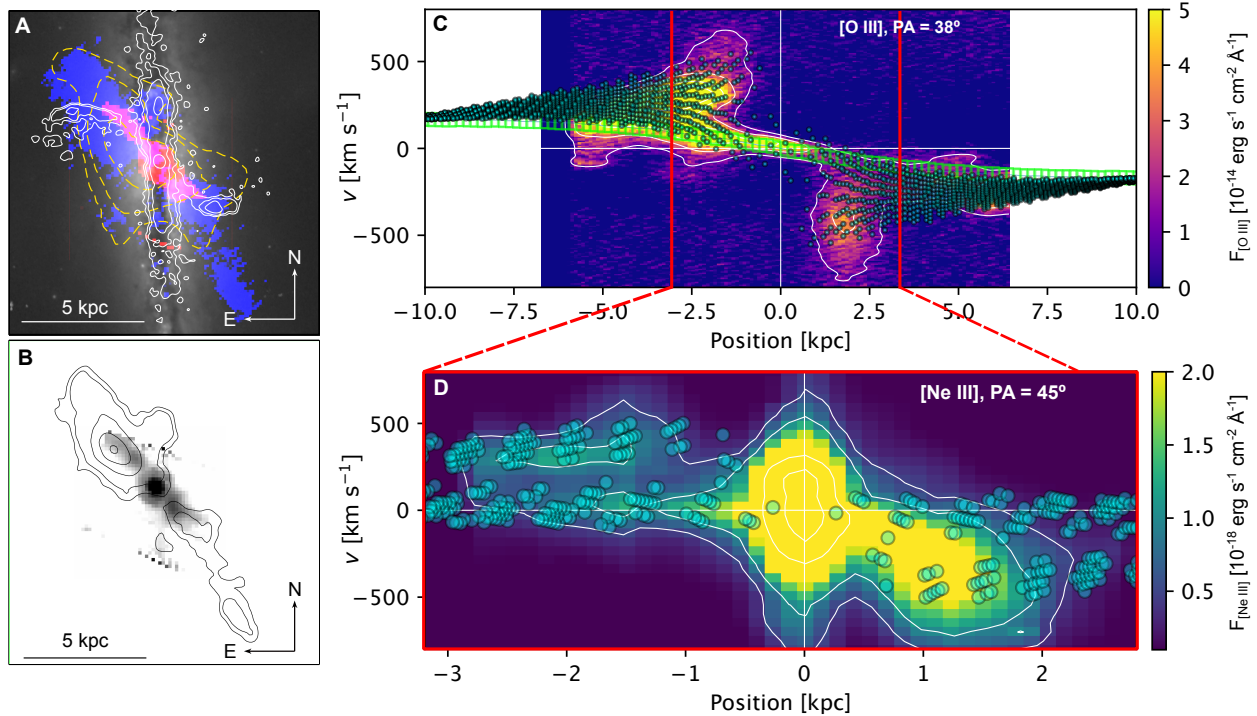

**Fig. S5**

**Multiwavelength observations and P-V diagrams for VV 340a.** Panel A the same as Fig. 1B, except the background *HST* broad band image is in grayscale and the blue channel is the high-resolution [O III] map from Keck KCWI in BH-S mode. Dashed yellow contours are the smoothed *Chandra* ACIS soft band X-ray image shown at the 4σ, 8σ, and 11σ levels. The X-ray emission follows the elongation of the coronal gas and continues linearly to the NE along the [O III] filament. Panel B is a greyscale image of the [Ne V] emission overlain with contours of [O III] emission at the 1.5σ, 2σ, 3σ, and 7σ levels from the high-resolution KCWI observation. The inner [O III] gas has a similar structure to the highly ionized gas traced by [Ne V]. Fig. 1B shows a lower-resolution [O III] image with a larger FOV but which does not resolve the structure in the central ~5 kpc region. Panels C Same as Fig. 3D, but for the higher-resolution KCWI BH-S data cube. (D) Same as Fig. 3E, but for the [Ne III] gas from the MIRI data cube. The model has a similar P-V distribution to the observed data in both panels.

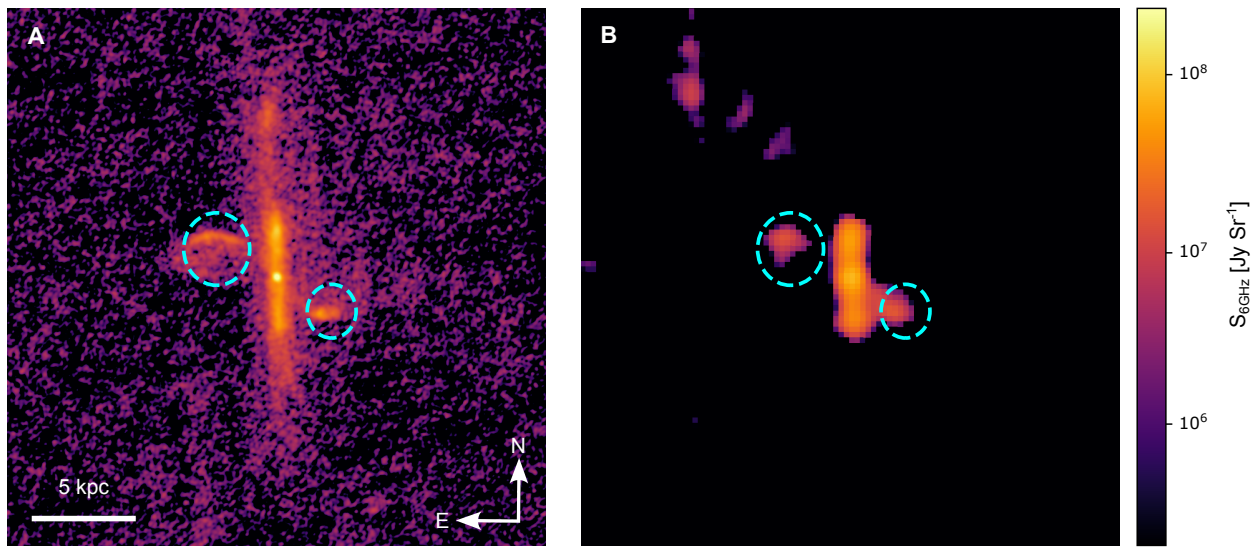

**Fig. S6**
**VLA radio images of VV 340a.** Panels show radio continuum images of VV 340a at 6 GHz (**A**) and 1.5 GHz (**B**). Maps are shown with identical surface brightness scales in units of Jy Sr$^{-1}$. Cyan dashed circles are 2.″7 and 2.″0 radius cylindrical apertures placed in the NE and SW, respectively, to measure the radio flux of the jet on either side of the galaxy.

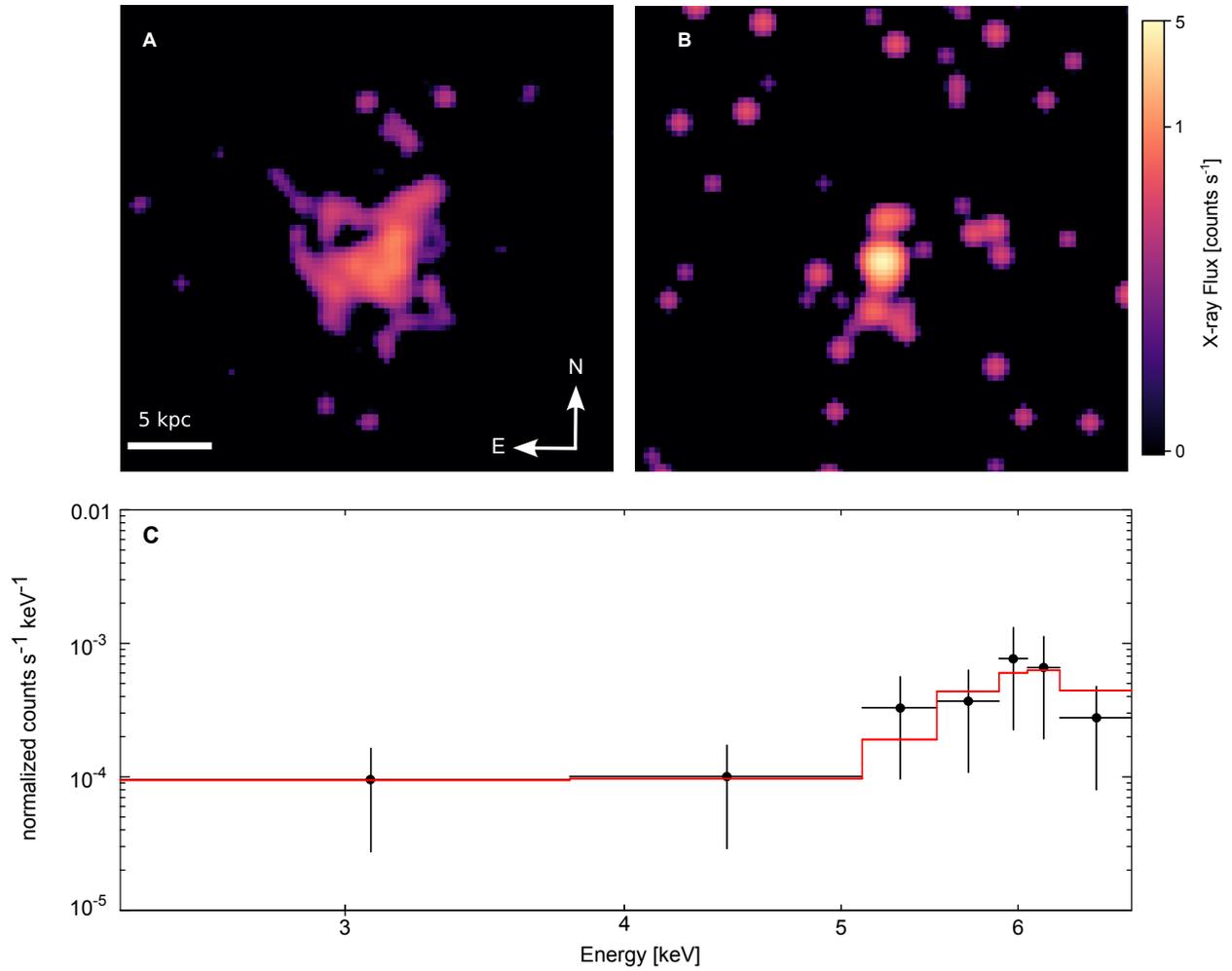

**Fig. S7**
***Chandra* soft and hard band X-ray images of VV 340 and X-ray energy spectrum extracted from the central 1″ region of VV 340a.** (**A**) Soft band (0.5-2 keV) image. (**B**) Hard band (2-7 keV) image. Both images have been smoothed with a Gaussian kernel with $r_{smooth}$ = 3 pixels and $\sigma_{smooth}$ = 1.5 pixels. (**C**) hard band energy spectrum extracted from the nuclear region of VV 340a (black filled circles with error bars) and best-fitting model spectrum (red) that assumes an intrinsic obscuring medium with a column density of $N_H = 3.7 \times 10^{23}$ cm$^{-2}$.

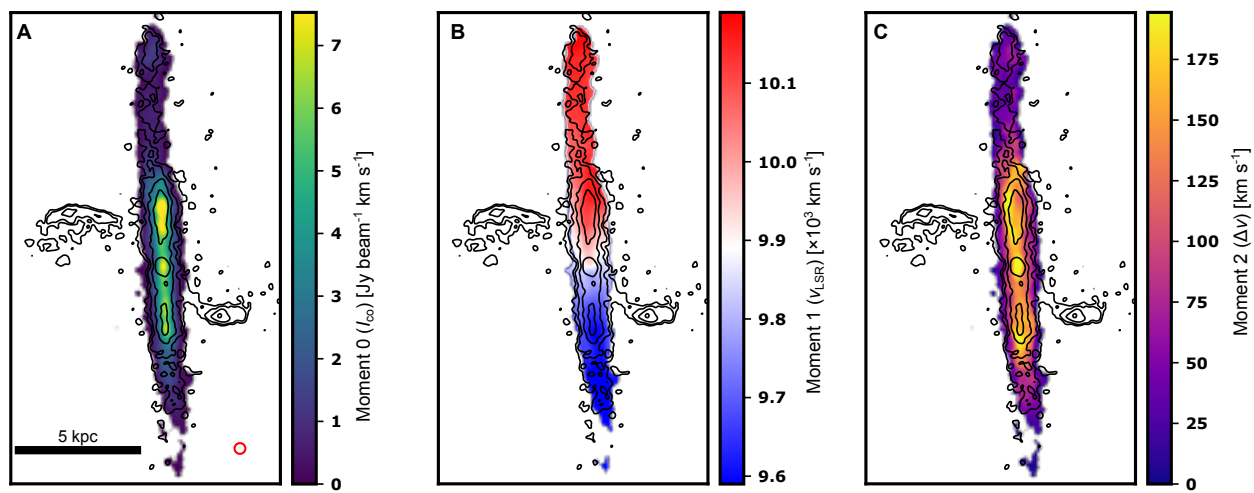

**Fig. S8**
**Distribution, velocity, and velocity dispersion of the cold molecular gas in VV 340a.** (A) the moment 0 (intensity), (B) moment 1 (intensity weighted mean velocity), and (C) moment 2 (velocity dispersion) maps of the CO (2-1) emission line from the ALMA observations. These trace the cold molecular gas in VV 340a. The intensity weighted mean velocity, $v_{LSR}$, is measured with respect to the local standard of rest and includes the systemic velocity of the galaxy. Black contours are the same VLA 6 GHz contours shown in Fig. 1. The FWHM of the primary beam (0.″4) is shown as the red circle in panel A. The molecular gas follows the expected distribution for a rotating disk, with no cold molecular gas component in the outflow.